\documentclass[11pt]{elsart}
\usepackage{graphicx}
\usepackage{theorem}
\usepackage{amssymb}

[1999/03/29 PSNFSS v.7.2
Times font as default roman
: S Rahtz]

\newcommand{\be}{\begin{equation}}
\newcommand{\ee}{\end{equation}}
\newcommand{\ba}{\begin{eqnarray}}
\newcommand{\ea}{\end{eqnarray}}
\newcommand{\E}{\rm E}

\theoremstyle{break}
\newtheorem{theorem}{Theorem}
\newtheorem{proposition}{Proposition}

\begin{document}

\begin{frontmatter}
\title{From Rational Bubbles to Crashes}

\author[Nice,UCLA]{D. Sornette} and
\author[Nice]{Y. Malevergne}

\address[Nice]{Laboratoire de Physique de la Mati\`ere Condens\'ee\\
CNRS UMR 6622 and
Universit\'e de Nice-Sophia Antipolis\\ 06108 Nice Cedex 2, France}
\address[UCLA]{Institute of Geophysics and Planetary Physics\\
and Department of Earth and Space Science\\
University of California, Los Angeles, California 90095, USA}
\address{email: Yannick.Malevergne@unice.fr and sornette@unice.fr\\
fax: (33) 4 92 07 67 54}

\begin{abstract}
We study and generalize in various ways the model of rational expectation (RE)
bubbles introduced by
Blanchard and Watson in the economic literature. Bubbles are argued
to be the equivalent of Goldstone modes of the fundamental rational pricing
equation, associated with the symmetry-breaking introduced by non-vanishing dividends.
Generalizing bubbles in terms of multiplicative stochastic maps,
we summarize the result of Lux and Sornette that the no-arbitrage condition
imposes that the tail of the return distribution is hyperbolic with an
exponent $\mu<1$. We then outline the main results of Malevergne and Sornette,
who extend the RE bubble model to arbitrary dimensions $d$:
a number $d$ of market time series are made linearly interdependent
via $d \times d$
stochastic coupling coefficients. We derive the no-arbitrage
condition in this context and, with the renewal theory for
products of random matrices applied to stochastic recurrence
equations, we extend the
theorem of Lux and Sornette to
demonstrate that the tails of the unconditional distributions
associated with such
$d$-dimensional bubble processes follow power laws, with the
same asymptotic tail exponent $\mu<1$ for all assets. The
distribution of price differences and of
returns is dominated by the same power-law over an extended range
of large returns.
Although power-law tails are a pervasive feature of empirical
data, the numerical value  $\mu<1$ is in disagreement with the usual
empirical estimates $\mu \approx 3$. We then discuss two extensions
(the crash hazard rate model and the non-stationary growth rate model)
of the RE bubble model that provide
two ways of reconciliation with the stylized facts of financial data.
\end{abstract}
\end{frontmatter}

\pagebreak

\section{The model of rational bubbles}

Blanchard \cite{Blanchard1} and Blanchard and Watson
\cite{Blanwat} originally introduced the model of
rational expectations (RE) bubbles to account for the possibility, often
discussed in the empirical literature and by practitioners, that observed
prices may deviate significantly and over extended time intervals
from fundamental prices.
While allowing for deviations from fundamental prices, rational bubbles
keep a fundamental anchor point of economic modelling, namely that bubbles
must obey the condition of rational expectations. In contrast,
recent works stress
that investors are not fully rational, or have at most bound
rationality, and that behavioral and psychological
mechanisms, such as herding, may be important in the shaping of market
prices \cite{Thaler,Shefrin,Shleifer}.
However, for fluid assets, dynamic investment strategies rarely perform
over simple buy-and-hold strategies
\cite{Malkiel}, in other words, the market is not far from being efficient
and little arbitrage opportunities exist as a result of the constant
search for gains by sophisticated investors. Here, we shall work within
the conditions of rational expectations and of no-arbitrage condition,
taken as useful approximations. Indeed, the rationality of both expectations
and behavior often
does not imply that the price of an asset be equal to its fundamental
value. In other words, there can be rational deviations of the price from
this value, called rational bubbles. A rational bubble can arise when
the actual
market price depends positively on its own expected rate of change, as
sometimes occurs in asset markets, which is the mechanism underlying the
models
of \cite{Blanchard1} and \cite{Blanwat}.

In order to avoid the
unrealistic picture of ever-increasing deviations from fundamental values,
Blanchard \cite{Blanwat} proposed a model with periodically
collapsing bubbles in
which the bubble component of the price follows an
exponential explosive path (the price being multiplied
by $a_t={\bar a}>1$) with probability $\pi$ and collapses to zero
(the price being multiplied
by $a_t=0$) with probability $1 - \pi$. It is clear that, in this model,
a bubble has an exponential
distribution of lifetimes with a finite average lifetime $\pi/(1-\pi)$.
Bubbles are thus transient phenomena. The condition
of rational expectations imposes that ${\bar a}=1/\delta$, where
$\delta$ is the discount factor.
In order to allow for the start of new bubbles after the collapse, a
stochastic
zero mean normally distributed component $b_t$ is added to the
systematic part of $X_t$.
This leads to the following dynamical equation
\be
X_{t+1} = a_t X_t + b_t,    \label{eq1}
\ee
where, as we said,
  $a_t={\bar a}$ with probability $\pi$ and $a_t=0$ with probability $1 -
\pi$. Both variables $a_t$ and $b_t$ do not depend on the process $X_t$.
There is a huge literature
on theoretical refinements of this model and on the empirical
detectability of RE bubbles in financial data (see
\cite{Camerer} and \cite{adamsz}, for surveys of this literature).
Model (\ref{eq1}) has also been explored in a large variety of contexts, for
instance in ARCH processes in econometry \cite{Haan},  1D random-field Ising
models \cite{Calan} using Mellin transforms, and more recently using
extremal properties of the $G -${\it
harmonic} functions on non-compact groups \cite{Solomon} and the Wiener-Hopf
technique \cite{Sorcont}. See also \cite{Sorknopoff} for a short review
of other domains of applications including
population dynamics with external sources, epidemics, immigration and
investment portfolios,
the internet, directed polymers in random media...

Large $|X_k|$ are generated by intermittent amplifications
resulting from the multiplication by several
successive values of $|a|$ larger than one. We now offer a simple
``mean-field'' type argument that clarifies the origin of the power
law fat tail.
Let us call $p_>$ the probability
that the absolute value of the multiplicative factor $a$
is found larger than $1$. The probability
to observe $n$ successive multiplicative factors $|a|$ larger than $1$ is
thus $p_>^n$. Let us call $|a_>|$ the average of $|a|$ conditionned
on being larger than $1$:
$|a_>|$ is thus the typical absolute value of the
  amplification factor. When $n$ successive multiplicative
factors occur with absolute values larger than $1$, they typically
lead to an amplification
of the amplitude of $X$
by $|a_>|^n$. Using the fact that the additive term $b_k$ ensures
that the amplitude
of $X_k$
remains of the order of the standard deviation or of other
measures of typical scales $\sigma_b$ of the distribution $P_b(b)$ when the
multiplicative factors $|a|$ are less than $1$,
this shows that a value of $X_k$ of the order of $|X| \approx
\sigma_b |a_>|^n$ occurs
with probability
\be
p_>^n = \exp \left( n \ln p_> \right) \approx \exp \left( \ln p_>
{\ln {|X| \over \sigma_b} \over \ln |a_>|} \right)= {1 \over
(|X|/\sigma_b)^{\mu}}~\,
\label{nckak}
\ee
with $\mu = \ln p_> / \ln |a_>|$, which can be rewritten as
$p_> |a_>|^{\mu}=1$. Note the similarity between this last ``mean-field''
equation and the exact solution (\ref{nfakka}) given below.
The power law distribution is thus the
result of an exponentially small probability of creating an exponentially
large value \cite{mybook}.
Expression (\ref{nckak}) does not provide a precise determination of the
exponent $\mu$, only an approximate one since we have used a kind of mean-field
argument in the definition of $|a_>|$.

In the next section, we recall how bubbles appear as
possible solutions of the fundamental pricing equation and play the role
of Goldstone modes of a price-symmetry broken by the dividend flow.
We then describe the Kesten generalization of rational bubbles
in terms of random multiplicative maps and present the fundamental
result \cite{Luxsor}  that the no-arbitrage condition leads to the
constraint that the exponent of the power law tail is less than $1$.
We then present an extension to arbitrary multidimensional random
multiplicative maps:
a number $d$ of
market time series are made linearly interdependent via $d \times d$
stochastic coupling
coefficients. We show that the no-arbitrage
condition imposes that the non-diagonal impacts of any asset $i$ on any
other asset $j \neq i$ has to vanish on average, i.e., must exhibit
random alternative regimes of reinforcement and contrarian feedbacks.
In contrast, the diagonal terms must be positive and equal on average
to the inverse
of the discount factor. Applying the results of renewal theory for
products
of random matrices to stochastic recurrence equations (SRE), we extend the
theorem of \cite{Luxsor} and
demonstrate that the tails of the unconditional distributions
associated with such
$d$-dimensional bubble processes follow power laws
(i.e., exhibit hyperbolic decline), with the
same asymptotic tail exponent $\mu<1$ for all assets. The
distribution of price differences and of
returns is dominated by the same power-law over an extended range
of large returns. In order to unlock the paradox, we
briefly discuss the crash hazard rate model \cite{JSL,JLS} and the
non-stationary growth model \cite{growthbubble}.
We conclude by proposing a link with the theory of speculative
pricing through a
spontaneous symmetry-breaking \cite{Sponsym}.

We should stress that, due to the no-arbitrage condition that forms
the backbone of our theoretical approach, correlations of returns
are vanishing. In addition, the multiplicative stochastic structure of the
models ensures the phenomenon of volatility clustering. These two
stylized facts, taken for granted in our present approach, will not be
discussed further.

\section{Rational bubbles of an isolated asset \cite{Luxsor}}

\subsection{Rational expectation bubble model and Goldstone modes}

We first briefly recall that pricing of an asset under
rational expectations theory is based on the two following hypothesis
: the rationnality of the agents and the ``no-free lunch'' condition.

Under the rationnal expectation condition, the best estimation of the
price $p_{t+1}$ of an asset at time $t+1$ viewed from time $t$ is
given by the expection of $p_{t+1}$ conditionned upon the knowledge of
the filtration $\{ {\mathcal F}_t \}$ (i.e. sum of all available information
accumulated) up to time $t$ : $E[p_{t+1}| {\mathcal F}_t]$.

The ``no-free lunch'' condition imposes that the expected returns of every assets
are all equal under a given probability measure ${\mathbb Q}$ equivalent
to the historical probability measure ${\mathbb P}$. In particular,
the expected return of each asset is equal to the return $r$ of the
risk-free asset (which is assumed to exist), and thus the probability measure ${\mathbb Q}$
is  named the {\it  risk neutral probability measure}.

Puting together these two conditions, we are led to the following
valuation formula for the price $p_t$~:
\be
\label{eqfundprice}
p_t = \delta \cdot \E_{\mathbb Q}[p_{t+1}|{\mathcal F}_t] + d_t ~ ~~~~~~~
\forall \{p_t\}_{t \geq 0}~,
\ee
where $d_t$ is an exogeneous ``dividend'', and $\delta = (1+r)^{-1}$ is the discount factor.
The first term in the r.h.s. quantifies the usual fact
that something tomorrow is less valuable than today by a factor called the
discount factor. Intuitively,
the second term, the dividend, is added to express the fact that
the expected price tomorrow has to be decreased by the dividend since the
value before giving the dividend incorporates it in the pricing.

The ``forward'' solution of (\ref{eqfundprice}) is well-known to be the fundamental price
\be
\label{eqnsolfund}
p_t^f = \sum_{i=0}^{+\infty} \delta^i \cdot \E_{\mathbb Q}[d_{t+i}|{\mathcal F}_t]~.
\ee
It is straightforward to check by replacement that the sum of the forward solution (\ref{eqnsolfund})
and of an arbitrary component $X_t$
\be
p_t  = p_t^f +  X_t~,
\label{jgjala}
\ee
where $X_t$ has to obey the single condition of being an arbitrary martingale:
\be
X_t = \delta \cdot \E_{\mathbb Q}[X_{t+1}|{\mathcal F}_t]    \label{aaajqjak}
\ee
is also the a solution of (\ref{eqfundprice}). In fact, it can be
shown \cite{GouLaMon}  that (\ref{jgjala}) is the general
solution of (\ref{eqfundprice}).

Here, it is important to note that, in the framework of the Blanchard
and Watson model, the speculative
bubbles appear as a natural consequence of the valuation formula
(\ref{eqfundprice}), i.e., of the no free-lunch condition and the rationnality of the agents.
Thus, the concept of bubbles is not an addition to the theory, as
sometimes believed, but is entirely embedded in it.

Notice also that the component $X_t$ in (\ref{jgjala}) plays a role
analogous to the Goldstone mode in nuclear, particle and
condensed-matter physics \cite{Goldstone1,Goldstone2}.
Goldstone modes are the zero-wavenumber zero-energy
modal fluctuations that attempt to restore a broken symmetry.
For instance, consider a Bloch wall between two semi-infinite
magnetic domains of opposite
spin directions selected by opposite magnetic field at boundaries far away.
At non-zero temperature, capillary waves are excited by thermal
fluctuations. The limit
of very long-wavelength capillary modes correspond to Goldstone modes that tend
to restore the translational symmetry broken by the presence of the Block wall
\cite{wave}.

In the present context, as shown in \cite{Sponsym}, the
\be
p \to -p ~~~~{\rm parity ~symmetry}    \label{fjallafc}
\ee
is broken by the ``external'' field embodied in the dividend flow $d_t$.
Indeed, as can be seen from (\ref{eqfundprice}) and its forward
solution (\ref{eqnsolfund}), the fundamental price is identically zero
in absence of dividends. Ref. \cite{Sponsym} has stressed the fact that
it makes perfect sense to think of {\it negative} prices.
For instance, we are ready to pay a
(positive) price for a commodity that we need or like. However, we
will not pay a positive
price to get something we dislike or which disturb us, such as garbage, waste,
broken and useless car, chemical and industrial hazards, etc.
Consider a chunk of waste. We will be ready to
buy it for a negative price, in order words, we are ready to take
the unwanted commodity if it comes with cash. Positive dividends imply
positive prices, negative dividends lead to negative prices. Negative dividends
correspond to the premium to pay to keep an asset for instance.
 From an economic view point, what makes a share of a company
desirable is its earnings,
that provide dividends, and its potential appreciation that give rise to
capital gains. As a consequence, in absence of dividends and of speculation,
the price of share must be nil and the symmetry (\ref{fjallafc}) holds.
The earnings leading to dividends $d$ thus act as a symmetry-breaking
``field'',
since a positive $d$ makes the share desirable and thus develop a
positive price.

It is now clear that the addition of the bubble $X_t$, which can be
anything but
for the martingale condition (\ref{aaajqjak}), is playing the role of
the Goldstone
modes restoring the broken symmetry: the bubble price can wander up
or down and, in the limit where it becomes very large in absolute value,
dominate over the fundamental price, restoring
the independence with respect to dividend.
Moreover, as in condensed-matter physics where the Goldstone mode
appears spontaneously since it has no energy cost,
the rationnal bubble itself can appear spontaneously with no dividend.

The ``bubble'' Goldstone mode turns out to be intimately related to the
``money'' Goldstone mode introduced by Bak et al. \cite{GoldstoneMoney}.
Ref.~\cite{GoldstoneMoney} introduces a
dynamical many-body theory of money, in which the value of
money in equilibrium is not fixed by the equations, and thus
obeys a continuous symmetry. The dynamics breaks this continuous
symmetry by fixating the value of
money at a level which depends on initial conditions. The
fluctuations around the equilibrium, for instance in the presence of
noise, are governed by the Goldstone modes
associated with the broken symmetry. In apparent contrast,
a bubble represents the arbitrary
deviation from fundamental valuation. Introducing money, a given
valuation or price is equivalent to a certain amount of money. 
A growing bubble thus corresponds to the same asset becoming equivalent to more and
more cash. Equivalently, from the point of view of the asset,
this can be seen as cash devaluation, i.e., inflation.
The ``bubble'' Goldstone mode and the ``money'' Goldstone mode are thus
two facets of the same fundamental phenomenon: they both are left unconstrained
by the valuation equations.

\subsection{The no-arbitrage condition and fat tails}

Following \cite{Luxsor}, we study the implications of the RE bubble models
for the unconditional distribution of prices, price
changes and returns resulting from a general discrete-time
formulation
extending (\ref{eq1}) by allowing the multiplicative factor
$a_t$ to take arbitrary values and be
i.i.d. random variables drawn from some non-degenerate
probability density function (pdf)
$P_a(a)$. The model can also be generalized by considering
non-normal realizations of $b_t$ with distribution
$P_b(b)$ with $\E_{\mathbb P}[b_t]=0$, where $\E_{\mathbb P}[\cdot]$ is
the unconditionnal expectation with respect to the probability measure ${\mathbb P}$.

Provided $\E_{\mathbb P}[\ln a] < 0$ (stationarity
condition) and if there is a number $\mu$ such that
$0 < \E_{\mathbb P}[|b|^{\mu}] < +\infty$, such that
\be
\E_{\mathbb P}[|a|^{\mu}] = 1    \label{nfakka}
\ee
and such that
$\E_{\mathbb P}[|a|^{\mu} \ln |a|] < +\infty$, then the tail of the distribution of
$X$ is asymptotically (for large $X$'s) a power law \cite{K73,Goldie}
\be
P_X(X) ~dX \approx {C  \over |X|^{1+\mu}}~dX~,   \label{fkaka}
\ee
with an exponent $\mu$ given by the real positive solution of
(\ref{nfakka}).

Rational expectations require in addition that
the bubble component of asset prices obeys the ``no free-lunch'' condition
\be
\delta \cdot \E_{\mathbb Q}[X_{t+1}|{\mathcal F}_t] = X_t    \label{bjqjak}
\ee
where $\delta<1$ is the discount factor and the expectation is taken
conditional on the knowledge of the filtration (information) until time $t$.
Condition (\ref{bjqjak})
with (\ref{eq1}) imposes first
\be
\E_{\mathbb Q}[a] = 1/\delta  >1~,
\ee
and then
\be
\E_{\mathbb P}[a]  >1~, \label{jnaflaq}
\ee
on the distribution of the multiplicative factors $a_t$.

\begin{figure}
\begin{center}
\includegraphics[width=14cm]{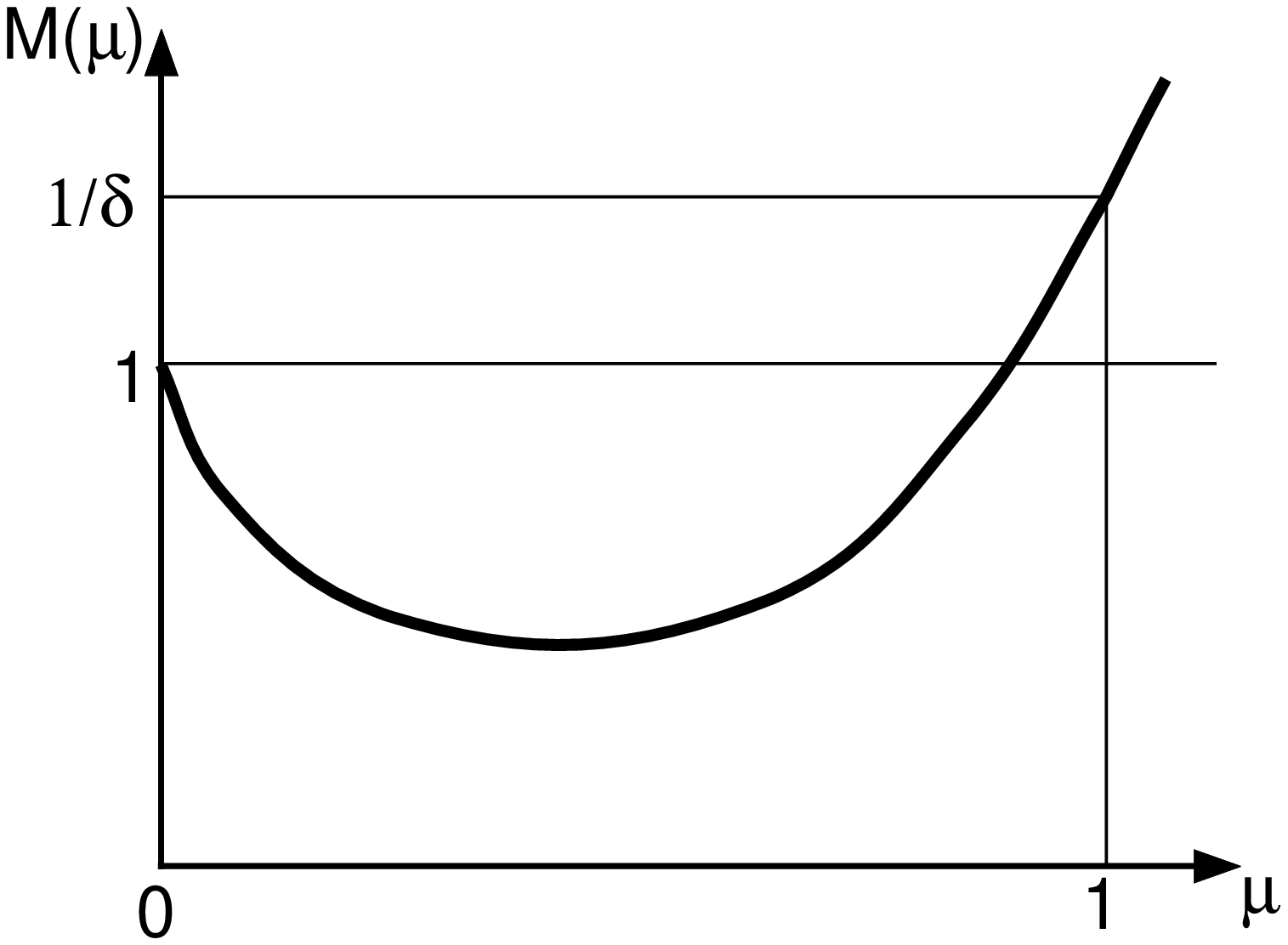}
\end{center}
\caption{Convexity of $M(\mu)$. This enforces that the exponent
solution of (\ref{nfakka})  is $\mu < 1$.}
\label{mdemu}
\end{figure}

\begin{figure}
\begin{center}
\includegraphics[width=14.cm]{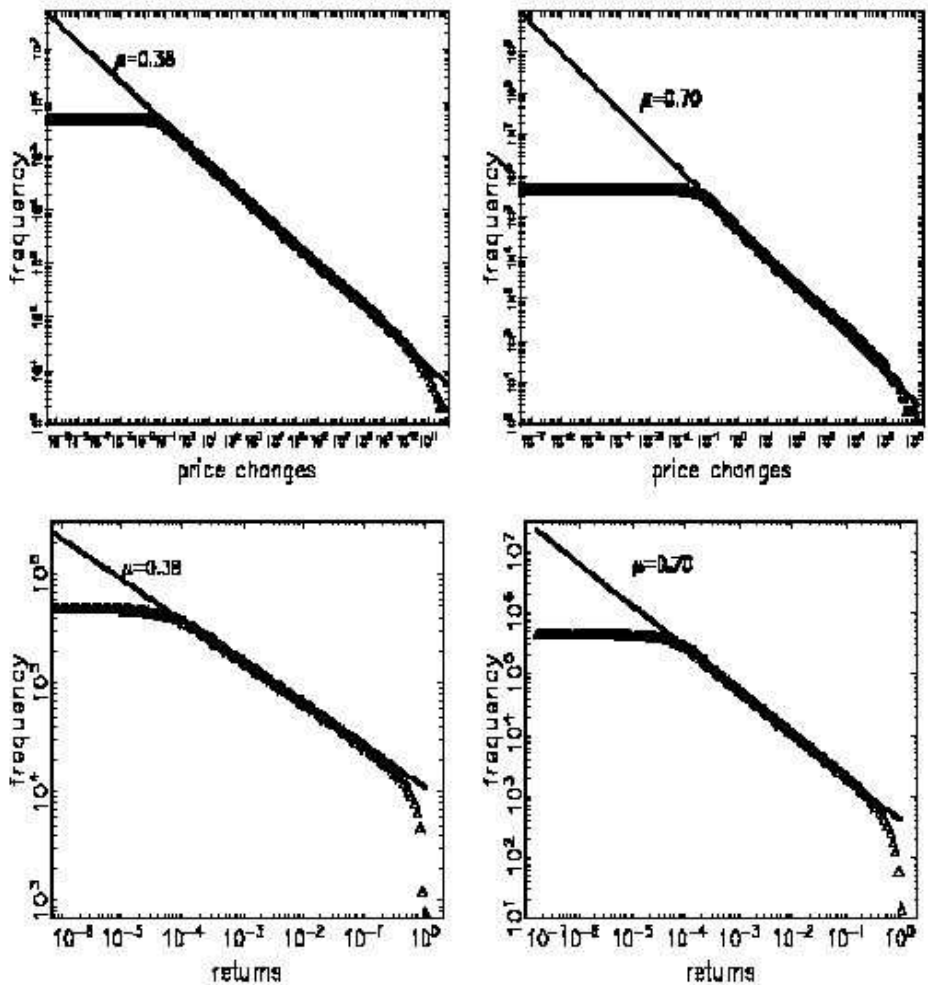}
\end{center}
\caption{Frequencies of large price changes and returns for selected bubble
processes: the plots exhibit the complement of the cumulative distribution from
simulations over $10^6$ time steps (triangles) and compares them with the
theoretically predicted tail behavior (straight line with slope
$\mu$). In all cases,
the scaling behavior is found to provide a good fit over an extended range of
magnitudes. However, while with price changes, the entire tail follows a Pareto
law with index $\mu$, returns are characterized by deviating behavior
at the highest
entries. Details on the underlying bubble processes are given in \cite{Luxsor}.
}
\label{figmu}
\end{figure}

Consider the function
\be
M(\mu) = \E_{\mathbb P}[a^{\mu}]~.
\ee
It has the following properties
\begin{enumerate}
\item $M(0)=1$ by definition,
\item $M'(0)=\E_{\mathbb P}[\ln a]<0$ from the stationarity condition,
\item $M''(\mu)=\E_{\mathbb P}[(\ln a)^2 |a|^\mu]>0$, by the positivity of the square,
\item $M(1)=1/\delta > 1$ by the no-arbitrage result (\ref{jnaflaq}).
\end{enumerate}
$M(\mu)$ is thus convex and is shown in figure \ref{mdemu}. This
demonstrate that $\mu < 1$ automatically (see \cite{Luxsor} for
a detailled demonstration).
It is easy to show \cite{Luxsor} that the distribution of price
differences has the same
power law tail
with the exponent $\mu<1$ and the distribution of returns is dominated by
the same power-law over an extended range of large returns \cite{Luxsor},
as shown in figure \ref{figmu}. Although power-law
tails are a pervasive feature of empirical data, these characterizations
are in strong disagreement with the usual empirical estimates which
find $\mu \approx 3$
\cite{devries,Lux,Pagan,Guillaume1,Gopikrishnan}.
Lux and Sornette \cite{Luxsor} concluded that
exogenous rational bubbles are thus hardly reconcilable with some of
the stylized facts of financial data at a very elementary level.

\section{Generalization of rational bubbles to arbitrary dimensions \cite{MalevergneSor}}

\subsection{Generalization to several coupled assets}

In reality, there is no such thing as an isolated asset. Stock markets
exhibit a variety of inter-dependences, based in
part on the mutual influences between the USA, European and
Japanese markets. In addition, individual stocks may be sensitive to
the behavior of the specific industry as a whole to which they belong
and to a few other
indicators, such as the main indices, interest rates and so on.
Mantegna et al. \cite{Mantegna1,Mantegna2}
have indeed shown the existence of a
hierarchical organization of stock interdepences. Furthermore,
bubbles often appear to be not isolated features of a set of markets.
For instance, ref. \cite{Flood} tested whether a
bubble simultaneously existed across the nations, such as
Germany, Poland, and Hungary, that experienced
hyperinflation in the early 1920s. Coordinated bubbles can sometimes
be detected. One of the most prominent example is found in the market
appreciations
observes in many of the world markets prior to the world market crash
in Oct. 1987 \cite{presidentcomit}.
Similar intermittent coordination of bubbles
have been detected among the significant bubbles followed
by large crashes or severe corrections in
Latin-American and Asian stock markets \cite{emer}.
It is therefore desirable to generalize the one-dimensional
RE bubble model (\ref{eq1}) to the multi-dimensional case.
One could also hope a priori that this generalization would modify the
result $\mu<1$ obtained in the one-dimensional case and allow for a
better adequation with empirical results. Indeed, 1d-systems
are well-known to exhibit anamalous properties often not shared
by higher dimensional systems. Here however, it turns out that
the same result $\mu<1$ holds, as we shall see.

In the case of several assets, rational pricing theory
again dictates that the fundamental price of each individual asset
is given by a formula like (\ref{eqfundprice}), where the specific
dividend flow of each asset is used, with the same discount factor.
The corresponding forward solution (\ref{eqnsolfund}) is again valid
for each asset. The general solution for each
asset is (\ref{jgjala}) with a bubble component $X_t$ different from
an asset to the next. The different bubble components can be coupled,
as we shall see, but they must each obey the martingale condition
(\ref{aaajqjak}), component by component. This imposes specific conditions
on the coupling terms, as we shall see.

Following this reasoning, we can therefore propose
the simplest generalization of a bubble into a ``two-dimensional''
bubble for two assets $\mathcal{X}$ and
$\mathcal{Y}$ with bubble prices
respectively equal to $X_t$ and $Y_t$ at time $t$. We express the
generalization of the Blanchard-Watson model as follows:
\be
X_{t+1} = a_t X_t + b_t Y_t + \eta_t    \label{eq11}
\ee
\be
Y_{t+1} = c_t X_t + d_t Y_k + \epsilon_t    \label{eq12}
\ee
where $a_t$, $b_t$, $c_t$ and $d_t$ are drawn from some
multivariate probability density function. The two additive noises
$\eta_t$ and $\epsilon_t$ are also drawn from some distribution function
with zero mean.
The diagonal case $b_t=c_t=0$ for all $t$ recovers the previous
one-dimensional case with two uncoupled bubbles, provided $\eta_t$ and
$\epsilon_t$ are independent.

Rational expectations require that $X_t$ and $Y_t$
obey both the ``no-free lunch'' condition (\ref{bjqjak}),
i.e., $\delta \cdot \E_{\mathbb Q}[X_{t+1}|{\mathcal F}_t] = X_t$ and $\delta \cdot
\E_{\mathbb Q}[Y_{t+1}|{\mathcal F}_t] =
Y_t$.
With (\ref{eq11},\ref{eq12}), this gives
\ba
\left(\E_{\mathbb Q}[a_t]-\delta^{-1}\right) X_t + \E_{\mathbb Q}[b_t] Y_t &=& 0~,   \label{hgaal}
\\
\E_{\mathbb Q}[c_t]X_t + \left(\E_{\mathbb Q}[d_t]-\delta^{-1}\right)  Y_t &=& 0~,  \label{fjkala}
\ea
where we have used that $\eta_t$ and $\epsilon_t$ are centered. The two equations
(\ref{hgaal},\ref{fjkala}) must be true for all times, i.e. for all
values of $X_t$ and $Y_t$ visited by the dynamics. This imposes
$\E_{\mathbb Q}[b_t]=\E_{\mathbb Q}[c_t]=0$ and $\E_{\mathbb Q}[a_t]=\E_{\mathbb Q}[d_t]=\delta^{-1}$.
We are going to retrieve this result more formally in the general case.

\subsection{General formulation}

A generalization to arbitrary dimensions leads to the
following stochastic random equation (SRE)
\be
\label{eq:sre}
{\bf X_t} = {\bf A_t X_{t-1}} + {\bf B_t}
\ee
where $({\bf X_t},{\bf B_t})$ are $d$-dimensional vectors. Each
component of ${\bf X_t}$
can be thought of as the price of an asset above its fundamental price.
The matrices $({\bf A_t})$ are
identically independent distributed $d \times d$-dimensional
stochastic matrices. We assume that ${\bf B_t}$ are
identically independent distributed random vectors and that $({\bf
X_t})$ is a causal
stationnary solution of (\ref{eq:sre}). Generalizations introducing
additional arbitrary linear
terms at larger time lags such as $X_{t-2}, ...$ can be treated
with slight modifications of our approach and yield the same
conclusions. We shall
thus confine our demonstration on the SRE of order $1$, keeping in mind
that
our results apply analogously to arbitrary orders of regressions.

In the following, we denote by $|\cdot|$ the Euclidean norm and by
$||\cdot||$
the corresponding norm for any
$d \times d$-matrix ${\bf A}$
\be
||{\bf A}|| = \sup_{|{\bf x}|=1} |{\bf Ax}| ~.
\ee
Technical details are given in \cite{MalevergneSor}.

\subsection{The no-free lunch condition}
\label{sec:nfl}

The valuation formula (\ref{eqfundprice}) and the martingale condition (\ref{aaajqjak}) given for a single asset easily
extends to a basket of assets. It is natural to assume that, for a given period $t$, the discount rate
$r_t(i)$, associated with asset $i$, are all the same. In frictionless markets, a
deviation for this
hypothesis would lead to arbitrage opportunities.
Furthermore, since the sequence of matrices $\{ \bf{A}_t \}$ is assumed to be i.i.d.
and therefore stationnary,
this implies that $\delta_t$ or $r_t$ must be constant and equal
respectively to $\delta$
and $r$.

Under those conditions, we have the following proposition:

\begin{proposition}
The stochastic process
\be
{\bf X_t} = {\bf A_t X_{t-1}} + {\bf B_t}
\ee
satisfies the no-arbitrage condition {\it if and only if}
\be
\E_{\mathbb Q}[{\bf A}] = \frac{1}{\delta} {\bf I_d}~.
\label{condnoarbi}
\ee
\end{proposition}

The proof is given in \cite{MalevergneSor} in which this
condition (\ref{condnoarbi}) is also shown to hold true under
the historical probability measure $\mathbb{P}$.

The condition (\ref{condnoarbi}) imposes some stringent constraints on
admissible matrices ${\bf A_t}$. Indeed, while ${\bf A_t}$ are not
diagonal in general,
their average must be diagonal. This implies that the off-diagonal terms of
the matrices ${\bf A_t}$ must take negative values, sufficiently often so
that their averages vanish. The off-diagonal coefficients quantify
the influence of other bubbles on a given one. The condition (\ref{condnoarbi})
thus means that
the average effect of other bubbles on any given one must vanish. It is
straightforward to check that, in this linear framework, this implies an
absence of correlation (but not an absence of dependence) between the different bubble components
$\E[X^{(k)} X^{(\ell)}]=0$ for any $k \neq \ell$.

In constrast, the diagonal elements of  ${\bf A_t}$ must be positive
in majority in order for $\E_{\mathbb P}[A_{ii}]={\delta^{(i)}}^{-1}$,
for all $i$'s, to hold true.
In fact, on economic grounds, we can exclude the cases
where the diagonal elements take negative values. Indeed, a negative value
of
$A_{ii}$ at a given time $t$ would imply that $X_t^{(i)}$ abruptly
change sign between
$t-1$ and $t$, what does not seem to be a reasonable financial process.

\subsection{Renewal theory for products of random matrices
\label{sectrenkes}}

In the following, we will consider that the random
$d \times d$ matrices ${\bf A}_t$ are invertible matrices with real
entries. We use the theorem 2.7 of Davis et al. \cite{DMB99}, which
synthetized  Kesten's theorems 3 and 4 in \cite{K73}, to the case of
real valued matrices. The proof of this theorem is given in \cite{LP83}.
We stress that the conditions listed below do not require the matrices
$({\bf A}_n)$ to be non-negative. Actually, we have seen that, in order
for the rational expectation condition not to lead to trivial results,
the off-diagonal coefficients of $({\bf A}_n)$ have to be negative with
sufficiently large probability such that their means vanish.

\begin{theorem}
Let $({\bf A}_n)$ be an i.i.d. sequence of matrices in
$GL_d(\mathbb{R})$ satisfying
the following set of conditions that we state in a heuristic manner
(see \cite{MalevergneSor} for technical details).
Provided that the following conditions hold,
\begin{itemize}
\item[H1 :] stationarity condition,

\item[H2 :] ergodicity,

\item[H3 :] intermittent amplification of the random matrices,

\item[H4 :] the fattailness of the distribution is not controlled
by that of the additive part $B_t$,
\end{itemize}

then, 
\begin{itemize}
\item there exists a unique solution $\mu \in (0, \mu_0]$ to the
equation \be
\label{eq:exp}
\lim_{n \rightarrow \infty} {1 \over n} \ln \E_{\mathbb P}\left[||{\bf
A}_1
\cdots {\bf A}_n||^{\mu}\right] = 0,
\ee
\item If $({\bf X}_n)$ is the stationary solution to the stochastic recurrence
equation in (\ref{eq:sre})
then ${\bf X}$ is regularly varying with index $\mu$. In other words,
the tail of the marginal distribution of each of the
components of the vector ${\bf X}$ is asymptotically a power law with
exponent $\mu$.
\end{itemize}
\end{theorem}

The equation (\ref{eq:exp}) determining the tail exponent $\mu$
reduces to (\ref{nfakka}) in the one-dimensional case, which is
simple to handle.
In the multi-dimensional case, the novel feature is the non-diagonal
nature of the multiplication of matrices which does not allow in general
for an explicit equation similar to (\ref{nfakka}).

It is important to stress that the tails of the distribution
of returns for all the components of the bubble
decrease with the same tail index $\mu$. This model thus provides a natural 
setting for rationalizing the well-documented empirical observation that
the exponent $\mu$ is found to be approximately the same for most assets. 
The constraint on its value discussed in the next paragraph does not 
diminishes the value of this remark, as explained in section 5.

\subsection{Constraint on the tail index}

The first conclusion of the theorem above shows that the tail index
$\mu$ of the process $({\bf X_t})$ is driven by the behavior of the
matrices (${\bf A_t}$). We will then state a proposition in which we
give a majoration of the tail index.

\begin{proposition}
A necessary condition to have $\mu >1$ is that the spectral radius
(largest eigenvalue) of $\E_{\mathbb P}[{\bf A}]$ be smaller than $1$~:
\be
\mu > 1 \Longrightarrow \rho(\E_{\mathbb P}[{\bf A}]) <1~.
\ee
\end{proposition}

The proof is given in \cite{MalevergneSor}.
This proposition, put together with
Proposition 1 above, allows us to derive our main result.
We have seen in section \ref{sec:nfl} from Proposition 1 that,
as a result of the no-arbitrage condition, the spectral radius of the
matrix $\E_{{\mathbb P}} [{\bf A}]$ is greater than $1$. As a consequence,
by application of the converse of Proposition 2, we find that the
tail index $\mu$ of the distribution of $({\bf X})$ is smaller
than $1$. This result does not rely on the diagonal property of the
matrices $\E_{\mathbb P}[{\bf A}_t]$ but only on the value of the spectral radius
imposed by the no-arbitrage condition.

This result generalizes to arbitrary $d$-dimensional processes the
result of \cite{Luxsor}. As a consequence, $d$-dimensional
rational expectation bubbles linking several assets suffer from the same
discrepancy compared to empirical data as the one-dimensional bubbles.
It would therefore appear that exogenous rational bubbles are hardly
reconcilable with some of the most
fundamental stylized facts of financial data at a very elementary level.

At this stage, we have to ask the question: what is wrong with the model
of rationnal bubbles?  Two alternative answers are explored below: 
either we believe in the standard
valuation formula and we are led to extend 
the restricted framework described by the Blanchard and Watson's model;
or we believe in the existence of the bubbles within their framework
and we have to generalize
the valuation formula. In the next two sections, we will discuss these two points of view.

\section{The crash hazard rate model \cite{JSL,JLS}}

In the stylised framework of {\it a purely speculative asset that pays no
dividends - i.e with zero fundamental price-} and in which we ignore information asymmetry
and the market-clearing condition, the price of the asset equals the price of the bubble
and the valuation formula (\ref{eqfundprice}) leads to the familiar martingale
hypothesis for the bubble price~:
\be \label{eq:martingale}
{\rm for~all}~~ t'>t ~~~~\delta_{t \to t'}{\rm E_{\mathbb Q}}[X(t') | {\mathcal F}_t] = X(t)~.
\ee
This equation is nothing but a generalisation of equation (\ref{aaajqjak}) to a continous time
formulation,  in which $\delta_{t \to t'}$ denotes the discount factor from time $t$ to time $t'$.

We consider a general bubble dynamics given by
\be \label{eq:crash}
dX = m(t)\,X(t)\,dt-\kappa X(t) dj~,
\ee
where $m(t)$ can be any nonlinear causal function of $X$ itself.
We add a jump process $j$ to capture the possibility that the bubble
exhibits a crash. $j$ is thus zero before the crash and one
afterwards. The random nature of the crash occurrence
is modeled by the cumulative distribution function $Q(t)$ of the time of the
crash, the probability density function
$q(t) = dQ/dt$ and the hazard rate $h(t) = q(t)/[1-Q(t)]$. The hazard
rate is the probability
per unit of time that the crash will happen in the next instant provided
it has not happened yet, i.e~:
\be
\label{eq:hr}
{\rm E_{\mathbb Q}}[dj | {\mathcal F}_t] = h(t) dt~.
\ee
Expression (\ref{eq:crash})
assumes that, during a crash, the bubble drops
by a fixed percentage $\kappa\in(0,1)$, say between $20$ and $30\%$ of the
bubble price.

Using $\E_{\mathbb Q}[X(t+dt) | {\mathcal F}_t ]=(1+rdt) X(t)$, where $r$ is the riskless
discount rate,
taking the expectation of (\ref{eq:crash}) conditioned on the filtration up to
time $t$ and using equation (\ref{eq:hr}), we get
\be
\E_{\mathbb Q}[dX | {\mathcal F}_t] = m(t) X(t)dt-\kappa X(t)h(t)dt = r X(t) dt~,
\ee
which yields\,:
\be
m(t)-r = \kappa h(t)~.
\label{hkllmqmlqm}
\ee
If the crash hazard rate $h(t)$ increases, the return $m(t)-r$
above the riskless interest rate
increases to compensate the traders
for the increasing risk. Reciprocally, if the dynamics of the bubble
shoots up, the rational expectation condition imposes an increasing crash risk
in order to ensure the absence of arbitrage opportunies: the risk-adjusted
return remains constant equal to the risk-free rate.
The corresponding equation for the bubble price, conditioned on the crash
not to have occurred, is\,:
\be
\label{eq:price1}
\log\left[\frac{X(t)}{X(t_0)}\right] = rt+ \kappa\int_{t_0}^t  h(t')dt'
\qquad\mbox{before the crash}.
\ee
The integral $\int_{t_0}^t h(t') dt'$ is the cumulative probability of a crash
until time $t$.
This gives the logarithm of the bubble price as the relevant observable. It has
successfully been applied to the 1929 and 1987 Wall Street
crashes up to about $7.5$ years prior to the crash \cite{SJ97,JLS}.

The higher the probability of a crash, the faster the bubble must increase
(conditional on having no crash) in order to satisfy the martingale condition.
Reciprocally, the higher the bubble, the more dangerous is the probability
of a looming crash.
Intuitively, investors must be compensated by a higher return
in order to be induced to hold an asset that might crash. This is the only
effect that this model captures. Note that the bubble dynamics can be
anything and the bubble can in particular be such that the distribution of
returns are fat tails with an exponent $\mu \approx 3$ without loss
of generality \cite{Jorgendidier}.

Ilinski \cite{Ilinski} raised the concern that the martingale condition
(\ref{eq:martingale})
leads to a model which ``assumes a zero return as the best prediction for
the market.'' He
continues\,: ``No need to say that this is not what one expects from
a perfect model of market bubble! Buying shares, traders expect the price
to rise and
it is reflected (or caused) by their prediction model. They support the bubble
and the bubble support them!''.
In other words, Ilinski \cite{Ilinski} criticises a key economic
element of the model
\cite{JSL,JLS} : market rationality. This point is captured by assuming
that the market level
is expected to stay constant (up to the riskless discount rate)
as written in equation (\ref{eq:martingale}).
Ilinski claims that this equation (\ref{eq:martingale})
is wrong because the market level does not stay constant in
a bubble\,: it rises, almost by definition.

This misunderstanding addresses a rather subtle point of the model and
stems from the difference between two different types of returns\,:
\begin{enumerate}
\item The unconditional return is indeed zero as seen from
(\ref{eq:martingale}) and reflects the fair game condition.

\item The conditional return, conditioned upon no crash occurring between
time $t$ and time $t'$, is non-zero and is given by equation
(\ref{hkllmqmlqm}). If the crash hazard rate is increasing
with time, the conditional return will be accelerating precisely because the
crash becomes more probable and the investors need to be remunerated for their
higher risk.
\end{enumerate}
Thus, the expectation which remains constant in equation (\ref{eq:martingale})
takes into account the probability
that the market {\em may} crash. Therefore, {\em conditionally}
on staying in the bubble (no crash yet), the market must rationally
rise to compensate buyers for having taken the risk that the market {\em
could} have crashed.

The market price reflects the equilibrium between the greed of buyers
who hope the bubble will inflate and the fear of sellers that it may crash.
A bubble that goes up is just one that could have crashed but did not.
The model \cite{JSL,JLS} is well summarised by borrowing the words of
another economist\,:
``(...) the higher probability of a crash leads to an acceleration
of [the market price] while the bubble lasts.''
Interestingly, this citation is culled from the very same article by
Blanchard \cite{Blanchard1} that Ilinski \cite{Ilinski} cites as an alternative
model more realistic than the model
\cite{JSL,JLS}. We see that this is in fact more
of an endorsement than an alternative.

A simple way to incorporate a different level of risk aversion into the model
\cite{JSL,JLS} is to say that
the probability of a crash in the next instant is perceived by traders
as being $K$ times bigger than it objectively is. This amounts to
multiplying our hazard rate $h(t)$ by $K$, and once again this makes
no substantive difference as long as $K$ is bounded away from zero and
infinity. Risk aversion is a central feature
of economic theory, and it is generally thought to be stable within a
reasonable range, associated with slow-moving secular trends such as
changes in education, social structures and technology.
Ilinski \cite{Ilinski} rightfully points out that risk perceptions are
constantly changing in the course of real-life bubbles, but wrongfully
claims that the model
\cite{JSL,JLS} violates this intuition. In this model, risk
perceptions do oscillate dramatically throughout the bubble, even
though subjective aversion to risk remains stable, simply because it is the
{\em objective degree of risk that the bubble may burst} that goes through
wild swings.
For these reasons, the criticisms put forth by Ilinski, far
from making a dent in the economic model \cite{JSL,JLS},
serve instead to show that it is robust, flexible and intuitive.

To summarize, the crash hazard rate model is such that
the price dynamics can be essentially arbitrary, and in particular such that
the corresponding returns exhibit a reasonable fat tail.  A jump
process for crashes
is added, with a crash hazard rate such that the rational
expectation condition is ensured.

\section{The non-stationary growth model \cite{growthbubble}}

In the previous section, we have presented a model which assumes that the
fundamental valuation formula remains valid and have generalized Blanchard 
and Watson's framework by reformulating the rational expection condition with
a jump crash process. We now consider the second view point
which consists in rejecting the validity of the valuation formula while keeping the
decomposition of the price of an asset into the sum of
a fundamental price and a bubble term. 
In this aim, we present a possible modification of the rational bubble
model of Blanchard and Watson, recently proposed in \cite{growthbubble},
which involves an average exponential growth of the fundamental price at
some return rate $r_f>0$ {\it larger than the discount rate}.

\subsection{Exponentially growing economy}

Recall that (\ref{jgjala}) shows that
the observable market price is
the sum of the bubble component $X_t$ and of a ``fundamental''
price $p_t^f$
\be
p_t = p_t^f + X_t~.   \label{jhgnakaa}
\ee
Thus, waiving off the valuation formula (\ref{eqfundprice}), let us assume that the fundamental price $p_t^f$ is growing
exponentially as
\be
p_t^f = p_0 e^{r_f t}  \label{hflklx}
\ee
at the rate $r_f$ and the
bubble price is following (\ref{eq1}).

Note that this formulation is compatible with the
standard valuation formula as long as $r_f < r$, provided the the cash-flow $d_t$ at time $t$
also growths with the same exponential rate $r_f$, i.e : $d_t=d_0 e^{r_f t}$. Indeed, the
 standard valuation formula then applies and leads to
\be
p_t^f = \sum_{k=0}^\infty \frac{d_t e^{k r_f}}{e^{kr}} \simeq \frac{d_0}{r-r_f} e^{r_f t}
\label{ghala}
\ee
up to the first order in $r_f-r$.  The discussion of the case $r_f > r$, for
which (\ref{ghala}) loses its meaning, is the subject 
of the sequel in which we follow \cite{Sponsym}.

Putting (\ref{eq1}) and (\ref{hflklx}) together with (\ref{jhgnakaa}), we obtain
\be
p_{t+1} = p_{t+1}^f + a_t X_t +  b_t = a_t p_t +
(e^{r_f} - a_t) p_t^f +  b_t~.  \label{jfkala}
\ee
Replacing $p_t^f$ in (\ref{jhgnakaa}) by $p_0 e^{r_f t}$ given in
(\ref{hflklx}) leads to
\be
p_t = e^{r_f t} \left(p_0 + \hat a_t \right)~.
\ee
where we have defined the ``reduced'' bubble price following
\be
\hat a_{t+1} = a_t e^{-r_f} \hat a_t + e^{-r_f} e^{-r_ft} b_t~.   \label{bgkaaAa}
\ee
Thus, if we allow the additive term $b_t$ in (\ref{eq1}) to also grow
exponentially as
\be
b_t = e^{r_ft} {\hat b}_t~,
\ee
where ${\hat b}_t$ is a stationary stochastic white noise process, we obtain
\be
\hat a_{t+1} = a_t e^{-r_f} \hat a_t + e^{-r_f} {\hat b_t}~,   \label{bgkaaA}
\ee
which is of the usual form. Intuitively,
the additive term represents the background of
``normal'' fluctuations around the fundamental price (\ref{hflklx}).
Their ``normal'' fluctuations have thus to grow with the same growth rate
in order to remain stationary in relative value.

In addition, replacing $p_t^f$ in (\ref{jfkala}) again by $p_0
e^{r_ft}$ leads to
\be
p_{t+1} = a_t p_t + e^{r_f t} [ p_0 (e^{r_f} - a_t) +  {\hat b}_t ]~.
\label{jfkaaaa}
\ee
The expression (\ref{jfkaaaa}) has the same form as (\ref{eq1}) with
a different additive term $[p_0 (e^r - a_t) +  {\hat b}_t ]$ replacing $b_t$.
The structure of this new additive term makes clear the origin of the
factor $e^{r_f t}$: as we said, it reflects
nothing but the average exponential
growth of the underlying economy. The contributions $b_t=e^{r_ft}
{\hat b}_t$ are then nothing
but the fluctuations around this average growth.

\subsection{The value of the tail exponent}

The condition $\E[\ln a] < r_f$ ensures that
$\E[\ln (a e^{-r_f})] < 0$ which is now the stationarity condition
for the process
$\hat a_t$ defined by (\ref{bgkaaA}). The conditions
$0 < \E[|e^{r_f} {\hat b}_t|^{\mu}] < +\infty$ (which is the same condition
$0 < \E[|b_t|^{\mu}] < +\infty$ as before) and
the solution of
\be
\E[|a e^{-r_f}|^{\mu}] = 1    \label{mnancac}
\ee
together with the constraint
$\E[|a e^{-r_f}|^{\mu} \ln |a e^{-r_f}|] < +\infty$ (which the same
as $\E[|a|^{\mu} \ln |a|] < +\infty$) leads to an asymptotic
power law distribution for the reduced price
variable $\hat a_t$ of the form $P_{\hat a}(\hat a) \approx C_{\hat a}/|\hat a|^{1+\mu}$, where
$\mu$ is the real positive solution of (\ref{mnancac}).
Note that the condition $\E[\ln (a e^{-r_f})] < 0$ which is
$\E[\ln (a)] < r_f$ now allows for positive average growth rate of the product
$a_t a_{t-1} a_{t-2} ... a_2 a_1 a_0$.

Consider
the illustrative case where the multiplicative factors $a_t$ are
distributed according to a log-normal distribution such that
$\E[\ln a] = \ln a_0$ (where $a_0$ is thus the most probable value taken
by $a_t$) and of variance $\sigma^2$. Then,
\be
\E[|a e^{-r_f}|^{\mu}] = \exp\left[ -r_f \mu + \mu \ln a_0 + \mu^2
{\sigma^2 \over 2}\right]~.
\label{gfjaa}
\ee
Equating (\ref{gfjaa}) to $1$ to get $\mu$ according to equation
(\ref{mnancac}) gives
\be
\mu = 2 {r_f - \ln a_0 \over \sigma^2} =
{r_f - \ln a_0 \over r - \ln a_0}  = 1 + {r_f - r
\over r - \ln a_0} ~.   \label{fjbalal}
\ee
We have used the notation $1/\delta=1+r$ for the discount
rate $r$
defined in terms of the discount factor $\delta$.
The second equality in (\ref{fjbalal}) uses
$\E[a] = a_0 e^{\sigma^2/2}$.

First, we retrieve the result \cite{Luxsor} that $\mu < 1$ for the
initial RE model
(\ref{eq1}) for which $r_f=0$ and $\ln a_0 < 0$.
However, as soon as $r_f > r \simeq -\ln \delta$, we get
\be
\mu >1 ~,   \label{jtlwlal}
\ee
and $\mu$ can take arbitrary values.
Technically, this results fundamentally from the structure of the
process in which
the additive noise grows exponentially
to mimick the growth of the bubble which alleviates the bound
$\mu<1$. Note that $r_f$ does not need to be large
for the result (\ref{jtlwlal}) to hold. Take for instance
an annualized discount rate $r^y=2\%$, an
annualized return $r^y_f=4\%$ and $a_0=1.0004$.
Expression (\ref{fjbalal}) predicts $\mu = 3$, which is compatible with
empirical data.

\subsection{Price returns}

The observable return is
$$
R_t = {p_{t+1} - p_t \over p_t} = {p_{t+1}^f - p_t^f +
X_{t+1} - X_t \over p_t^f + X_t} 
$$
\be
= \chi_t
\left({p_{t+1}^f - p_t^f \over p_t^f} + {X_{t+1} - X_t \over p_t^f}\right)
= \chi_t \left( r_f + {\hat a_{t+1} - \hat a_t \over p_0} \right)~,
  \label{jfalala}
\ee
where
\be
\chi_t = {p_t^f \over p_t^f + X_t} = {1 \over 1 + (\hat a_t/p_0)}~.
\label{jfoqloaq}
\ee
In order to derive the last equality in the right-hand-side of (\ref{jfalala}),
we have used the definition of the return of the fundamental price (neglecting
the small second order difference between $e^{r_f}-1$ and $r_f$).
Expression (\ref{jfalala}) shows that the distribution of returns $R_t$
of the observable prices is the same as that of the product of the random
variable $\chi_t$ by $r_f + (\hat a_{t+1} - \hat a_t)/p_0$. Now, the tail of the
distribution of $r_f + (\hat a_{t+1} - \hat a_t)/p_0$ is the same as the tail
of the distribution
of $\hat a_{t+1} - \hat a_t$, which is a power law with exponent $\mu$ solution
of (\ref{mnancac}), as shown rigorously in \cite{Luxsor}.

It remains to show that the product of this variable $r_f + (\hat a_{t+1}
- \hat a_t)/p_0$
by $\chi_t$ has the same tail behavior as $r_f + (\hat a_{t+1} - \hat a_t)/p_0$ itself.
If $r_f + (\hat a_{t+1} - \hat a_t)/p_0$ and $\chi_t$ were independent, this would follow
from results in \cite{Breiman} who demonstrates that for
two independent random variables $\phi$ and $\chi$ with
Proba$(|\phi|>x)\approx c x^{-\kappa}$  and
E$[\chi^{\kappa+\epsilon}]<\infty$ for some $\epsilon > 0$, the
random product $\phi \chi$ obeys Proba$(|\phi \chi|>x)\approx
{\rm E}[\chi^{\kappa}] x^{-\kappa}$.

However, $r_f + (\hat a_{t+1} - \hat a_t)/p_0$ and $\chi_t$ are not independent
as both contain
a contribution from the same term $\hat a_t$. However, when $\hat a_t<<p_0$,
$\chi_t$ is close to $1$ and the previous result should hold. The
impact of $\hat a_t$ in $\chi$ becomes important when $\hat a_t$ becomes
comparable to $p_0$.

It is then convenient to rewrite
(\ref{jfalala}) using (\ref{jfoqloaq}) as
\be
R_t = {r_f \over 1+ (\hat a_t/p_0)} + {\hat a_{t+1} - A_t \over p_0 + \hat a_t}
={r_f \over 1+ (\hat a_t/p_0)} + {(a_t e^{-r_f} -1) \hat a_t + e^{-r_f} b_t
\over p_0 + \hat a_t}~.
\label{jfjncal}
\ee
We can thus distinguish two regimes:
\begin{itemize}
\item for not too large values of the reduced bubble term $\hat a_t$, specifically
for $\hat a_t < p_0$, the denominator $p_0+ \hat a_t$
changes more slowly than the numerator of the second term,
so that the distribution of returns will
be dominated by the variations of this numerator
$(a_t e^{-r_f} -1) \hat a_t + e^{-r_f} b_t$ and, hence, will follow
approximately the same power-law as for $\hat a_t$, according to the results of
\cite{Breiman}.

\item For large bubbles, $\hat a_t$ of the order of or greater than $p_0$,
the situation changes, however: from (\ref{jfjncal}), we
see that when the reduced bubble term $\hat a_t$ increases without bound, the first
term $r_f/(1+ (\hat a_t/p_0))$ goes to $0$ while the second term becomes
asymptotically
$a_t e^{-r} -1$. This leads to the existence of an absolute upper bound for the
absolute value of the returns.
\end{itemize}

\begin{figure}
\begin{center}
\includegraphics[height=15cm,width=16cm]{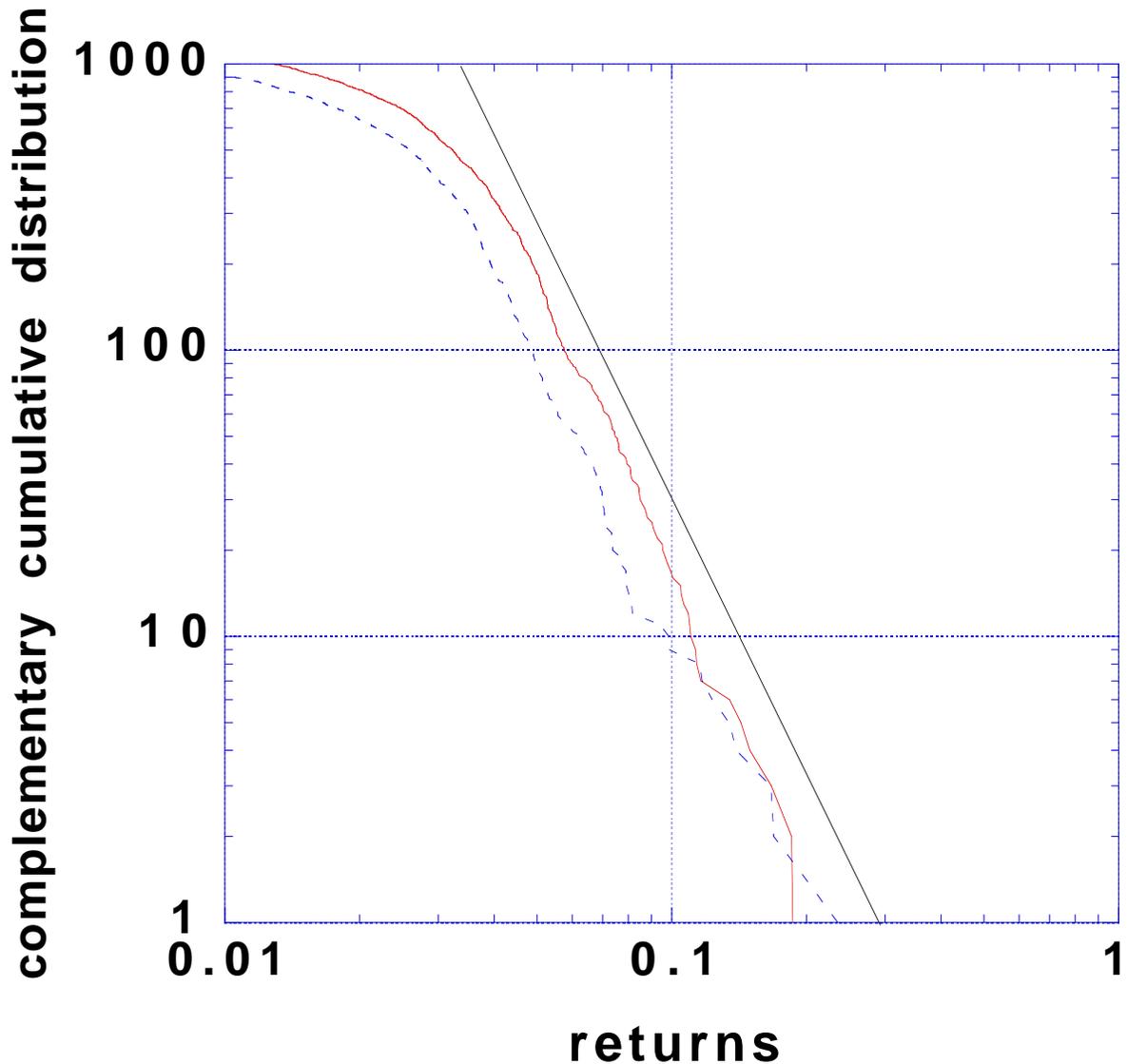}
\caption{\label{fig3} Double logarithmic scale representation of the
complementary cumulative distribution of the
``monthly'' returns $R_t$ defined in (\ref{jfalala}) of the synthetic total
price sum of the exponential growing fundamental price and the bubble price.
The continuous (resp. dashed) line corresponds to the positive (resp.
negative) returns.
The distribution is
well-described by an asymptotic power law with an exponent in agreement
with the prediction
$\mu \approx 3.3$ given by the equations (\ref{mnancac})
and (\ref{fjbalal}) and shown as the straight
line. The small differences between the predicted slope and the
numerically generated
ones are within the error bar of $\pm 0.3$ obtained from a standard maximum
likelihood Hill estimation. From \cite{growthbubble}.
}
\end{center}
\end{figure}

To summarize, we expect
that the distribution of returns will therefore follow a power-law with the
same exponent $\mu$ as for $\hat a_t$, but with a finite cut-off (see
\cite{growthbubble} for details).
This is validated by numerical simulations
shown in figure \ref{fig3} taken from \cite{growthbubble}.

Thus, when the price fluctuations
associated with bubbles on average grow with the mean market return
$r_f$, we find that the exponent of the power
law tail of the returns is no more bounded by $1$ as soon as $r_f$
is larger than the discount rate $r$ and can take essentially
arbitrary values. It is remarkable that this condition
$r_f > r$ corresponds to the paradoxical and unsolved regime
in fundamental valuation theory where the forward valuation solution
(\ref{eqnsolfund})
loses its meaning, as discussed in \cite{Sponsym}. In analogy with the theory
of bifurcations and their normal forms, ref.\cite{Sponsym} proposed
that this regime
might be associated with a spontaneous symmetry breaking phase corresponding
to a spontaneous valuation in absence of dividends by pure
speculative imitative
processes.

\section*{Conclusion}

Despite its elegant formulation of the bubble phenomenon,
the Blanchard and Watson's model suffers from a lethal discrepancy: it does not seem to
comply with the empirical data, i.e., it cannot generate power law tails whose exponent is
greater than $1$, in disagreement with the empirical tail index found around $3$.
We have summarized the demonstration that this result holds true both for a bubble defined for
a single asset as well as for bubbles on any set of coupled assets, as long as the 
rational expectation condition holds.

In order to reconcile the theory with the empirical facts on the tails of the
distributions of returns, two alternative
models have been presented. The ``crash hazard rate'' model extends the formulation
of Blanchard and Watson by replacing the linear stochastic
bubble price equation by an arbitrary dynamics solely constrained by the no-arbitrage
condition made to hold with the introduction of a jump process.
The ``growth rate model'' departs more audaciously from standard economic models
since it discards one of the pillars of the standard valuation theory, but putting
itself firmly in the regime $r_f>r$ for which the fundamental valuation formula
breaks down. In addition to allowing for correct values of the tail exponent, its
provides a generalization of the fundamental valuation formula by providing 
an understanding of its breakdown as deeply associated with a spontaneous 
breaking of the price symmetry. Its implementation for multi-dimensional bubbles
is straightforward and the results obtained in section 5 carry over naturally
in this case.
This provides an explanation for why the tail index $\mu$ seems to be the same for any
group af assets as observed empirically. This work begs for the introduction of 
a generalized field theory which would be able to capture the spontaneous 
breaking of symmetry, recover the fundamental valuation formula in the normal
economic case $r_f < r$ and extend it to the still unexplored regime $r_f>r$.

\vspace{0.5cm}

{\bf Acknowledgement}: We acknowledge helpful discussions and exchanges with
J.V. Andersen, A. Johansen, J.P.
Laurent, O. Ledoit,T. Lux, V. Pisarenko and M. Taqqu and thank T.
Mikosch for providing
access to \cite{LP83}.

\pagebreak

\pagebreak

\end{document}